\documentstyle[buckow,epsfig]{article}

\def\s{\sigma}

\def\g{\gamma}

\newcommand{\identita}{ \,  \raisebox{+0.14em}{{\hbox{{\rm \scriptsize ]}} \raisebox{-0.2em}{\kern-.8em\hbox{1}}}} \, }

\newcommand{\bbox}{\lower.2ex\hbox{$\Box$}}

\newcommand{\bde}{\begin{description}}
\newcommand{\ede}{\end{description}}

\renewcommand{\a}{\alpha}
\renewcommand{\b}{\beta}

\newcommand{\pa}{\partial}
\newcommand{\G}{\Gamma}

\newcommand{\e}{\epsilon}

\newcommand{\m}{\mu}

\newcommand{\n}{\nu}

\renewcommand{\t}{\tau}

\newcommand{\Db}{\bar{D}}
\newcommand{\Wb}{\bar{W}}

\newcommand{\ad}{{\dot{\alpha}}}
\newcommand{\bd}{{\dot{\beta}}}
\newcommand{\gd}{{\dot{\gamma}}}

\newcommand{\DC}{\nabla}

\newcommand{\Tra}{{\rm{Tr}}_{Ad}}

\def\beq{\begin{equation}}
\def\eeq{\end{equation}}

\begin {document}
\def\titleline{
Nonabelian Born-Infeld from Super-Yang-Mills effective action
}
\def\authors{
A.Refolli\1ad, A.Santambrogio \2ad, N.Terzi \3ad and D.Zanon \4ad }
\def\addresses{
Dipartimento di Fisica dell'Universit\`a di Milano and \\
INFN, Sezione di Milano, Via Celoria 16,\\
20133 Milano, Italy\\
\1ad
{\tt Andrea.Refolli@mi.infn.it} \\
\2ad
{\tt Alberto.Santambrogio@mi.infn.it}\\
\3ad
{\tt Niccolo.Terzi@mi.infn.it}\\
\4ad
{\tt Daniela.Zanon@mi.infn.it}\\
}
\def\abstracttext{
We discuss recent results on one-loop contributions to the effective action  
in ${\cal N}=4$ supersymmetric Yang-Mills theory in four dimensions. 
Contributions with five external vector fields are compared with corresponding
ones in open superstring theory in order to understand the relation with
the $F^5 $ terms that appear in the nonabelian generalization of the 
Born-Infeld action. }
\large
\makefront

\section{Introduction}
This talk is based on the paper \cite{noi}.
Scattering amplitudes of massless modes in string theories can be described in terms of effective  Lagrangians. For the open string case the abelian Born-Infeld action represents a remarkable example of an effective field theory which contains string corrections to all orders in $\a'$ \cite{FT}. The contributions can be summed up to all orders
in the case of a constant, abelian field strength. As soon as these two 
conditions are relaxed the problem becomes complicated and a complete action for such fields has not been obtained. A non-abelian generalization of the Born-Infeld action can be defined as suggested in \cite{Tseytlin1} using a symmetrized trace operation over the gauge group matrices. However there are indications that this prescription might not be sufficient to include all the contributions that one would obtain from an open string approach \cite{HT,BRS}.

The best available way to  construct the effective action for the non-abelian, non constant curvature case, is to proceed order by order. One has to compute corrections with increasing number of derivatives and of field strengths. A well established result is up to the order $\a'^2$ \cite{GW}.
As soon as one focuses on higher orders the calculations become difficult. The inclusion of supersymmetry seems to be quite useful to gain insights and it might set enough constraints to fix uniquely the form of the action. Several attempts in different directions are under consideration \cite{BRS,molti}.

In our paper \cite{noi} we have attacked the problem as follows:
since we want to make contact with the ten-dimensional open superstring we 
have considered its four-dimensional field theory limit, i.e. the ${\cal N}=4$ 
supersymmetric Yang-Mills theory. Then we have computed at one-loop 
perturbatively: the idea is that in this way we construct an effective action 
which is supersymmetric and generalizes the Yang-Mills theory. If 
supersymmetry determines the form of the allowed deformations 
\cite{uniqueness} it should correspond to the non-abelian Born-Infeld theory. 
However the crucial open question is: to what extent is supersymmetry enough to
fix the form of higher order corrections?


We present the calculation of the four- and five-point functions, and of the 
part of the six-point function which is needed for the covariantization of the
lower order results. These computations allowed us to evaluate the complete
gauge invariant structures for the  $F^4$ and the $\nabla\nabla F^4$, $F^5$ 
contributions. Since the symmetric trace prescription would rule out $F^5$
terms, the nonvanishing result that we find suggests a richer form for 
the non-abelian Born-Infeld action. In fact the trace operation on the gauge 
group indices receives contributions from the symmetric as well as the 
antisymmetric part. 

In the next section we briefly recall the ${\cal N}=1$ superspace formulation 
of the ${\cal N}=4$  Yang-Mills action and describe the main ingredients that 
enter the quantization via the background field approach.
In section 3 we present the one-loop amplitude up to order $\frac{1}{M^6}$
in the low-energy expansion. We have extracted from the superfield result its 
component content and in particular have studied the bosonic contributions 
which contain the field strengths $F_{\m\n}$. 
We then compare our results with corresponding ones from open superstring 
theory \cite{kitazawa} and with other results recently obtained with different
techniques in \cite{Bil,KS}.

\section{The ${\cal N}=4$  Yang-Mills action and its quantization}

The ${\cal N}=4$ supersymmetric Yang-Mills  classical action
written in terms of ${\cal N}=1$ superfields (we use the
notations and conventions adopted in \cite{superspace}) is given by
\begin{eqnarray}
&&S= \frac{1}{g^2}~{\rm Tr} \left( \int~ d^4x~d^4\theta~ e^{-V}
\bar{\Phi}_i e^{V} \Phi^i +\frac{1}{2} \int ~d^4x~d^2\theta~ W^\a W_\a
+\frac{1}{2} \int ~d^4x~d^2\bar{\theta}~ \bar{W}^\ad \bar{W}_\ad\right.
\nonumber\\
&&\left.~~~~~~~~~~~~~+\frac{1}{3!} \int ~d^4x~d^2\theta~ i\e_{ijk}
 \Phi^i
[\Phi^j,\Phi^k] + \frac{1}{3!}\int ~d^4x~d^2\bar{\theta}~ i\e^{ijk} 
\bar{\Phi}_i
[\bar{\Phi}_j,\bar{\Phi}_k] \right)
\label{N4SYMaction}
\end{eqnarray}
where the $\Phi^i$ with $i=1,2,3$ are three chiral
superfields, and the $W^\a= i\bar{D}^2(e^{-V}D^\a e^V)$ are the gauge
superfield strengths. All the fields are Lie-algebra valued, e.g.
$\Phi^i=\Phi^i_a T^a$, in the adjoint representation of the gauge group, with $[T_a,T_b] = i f_{abc} T_c$. 

Since we are interested in computing field strength corrections to the Born-Infeld action, we will consider amplitudes with vector fields as external background. Moreover we will extract from them only the gauge field bosonic components using the relations 
\begin{equation}
D_{(\a}W_{\b)}|_{\theta=0}=f_{\a\b}=\frac{1}{2} (\sigma_{\m\n})_{\a\b}F^{\m\n}~~~~~~~\Db_{(\ad}\Wb_{\bd)}|_{\theta=0}=\bar{f}_{\ad\bd}=-\frac{1}{2} (\bar{\sigma}_{\m\n})_{\ad\bd}F^{\m\n}
\label{component}
\end{equation}

We have used the background field method \cite{superspace}; after gauge-fixing the quadratic quantum $V$-action becomes
\begin{eqnarray}
&&S\rightarrow -\frac{1}{2g^2} {\rm Tr} \int
d^4x~d^4\theta~V\left[\Box-i {\bf \G}^{\g\gd}\pa_{\g\gd} -\frac{i}{2}
(\pa^{\g\gd}{\bf \G}_{\g\gd})-\frac{1}{2}{\bf \G}^{\g\gd}{\bf \G}_{\g\gd}\right.
\nonumber\\
&&~~~~~~~~~~~~~~~~~~~~~~~~~~~~~~~~~\left.-i {\bf W}^\a(D_\a-i{\bf\G}_\a)-i\bar{\bf W}^\ad
(\bar{D}_\ad-i{\bf\G}_\ad)\right]V
\label{actionquadratic}
\end{eqnarray}
The ${\cal N}=4$ theory is particularly simple since the loops with the three chiral matter fields are cancelled by the three ghosts. 

A one-loop $n$-point amplitude will contain $n$ external background fields from the interactions in (\ref{actionquadratic}), with at least two ${\bf W}$ and two $\bar{\bf W}$ in order to complete the $D$-algebra.
We set the external fields on-shell, i.e. we freely use the equations of motion
$\nabla^\a {\bf W}_\a=0$ and $\bar{\nabla}^\ad {\bf \Wb}_\ad=0$\\
We obtain the structures we are looking for if our one-loop diagram has produced products of fields $D_{(\a}{\bf W}_{\b)}$ and their hermitian conjugate ones $\Db_{(\ad}{\bf \Wb}_{\bd)}$.\\
In addition we have to deal with a loop-momentum integral which contains $n$ scalar propagators and momentum factors directly from the vertices in (\ref{actionquadratic}) and/or from commutators of spinor derivatives produced while performing the $D$-algebra
\begin{equation}
I_n=\int {\rm d}^4 k \frac{h_n(k,p_i)}{k^2 (k+p_2)^2(k+p_2+p_3)^2\dots(k+p_2+\dots+p_n)^2}
\label{npointint}
\end{equation}
We can rewrite (\ref{npointint}) as an infinite series of local terms in a low-energy expansion with higher derivatives by introducing an IR mass $M$ and expanding the propagators keeping the external momenta small as compared to $M$ (see \cite{noi} for more details). 
\section{${\cal N}=4$ Yang-Mills at one loop up to order $\frac{1}{M^6}$}
The four-point function corresponds to box-type diagrams as the one shown 
in Fig.1,
and gives rise to a momentum integral
\begin{eqnarray}
I_4 &=& \int {\rm d}^4 k \frac{1}{k^2 (k+p_2)^2(k+p_2+p_3)^2(k+p_2+p_3+p_4)^2} 
\label{box}
\end{eqnarray}
and to a background field dependence
\begin{equation}
\int d^4\theta ~\Tra({\bf W}^{\a}(p_1){\bf W}_{\a}(p_2){\bf \Wb}^{\ad}(p_3){\bf \Wb}_{\ad}(p_4)) 
\label{4background}
\end{equation}
where in general we have defined $\Tra(AB\cdots ) \equiv A_a B_b \cdots  g(a,b,\ldots)$, with the group colour factor $g(a_1,a_2,\ldots,a_n) = f_{x_1 a_1 x_2}f_{x_2 a_2 x_3}\cdots f_{x_n a_n x_1}$.\\ 
\begin{figure}[htb] \begin{center}
\mbox{\epsfig{figure={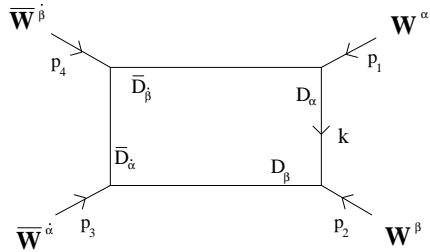},
       width=0.35 \linewidth}}
\caption{Four-point amplitude}
\label{Box} \end{center}
\end{figure} 
We are interested in rewriting the superspace result in components and in particular we want to study the bosonic gauge field contributions. 
If we do so we find that at the leading order in the local expansion the planar sector (large $N$ limit) reproduces the symmetrized trace structure \cite{Tseytlin1}
\begin{eqnarray}
 \G_4&\Rightarrow &{\rm Tr}\Big( T^{a_1}T^{a_2}T^{a_3}T^{a_4}+ {\rm permutations}\Big)\left\{(F^4)_{a_1a_2a_3a_4} -\frac{1}{4}(F^2)_{a_1a_2}(F^2)_{a_3a_4} \right\} \nonumber \\
& \equiv & STr\left(F^4 -\frac{1}{4}(F^2)^2 \right)
\label{4symmetric}
\end{eqnarray}
The subleading contribution corresponds to terms with two derivatives:
\begin{eqnarray} \label{eq:p^2}
\G_4^{(der)}&=& -\frac{1}{40}\left(\frac{\pi^2}{12}\frac{1}{M^6}\right)\left[10 (p_2)^2 + 8 p_2 \cdot p_3 + 2 p_2 \cdot p_4  \right]~g(a_1,a_2,a_3,a_4) \nonumber \\
&& \frac{1}{8}\left\{\left[ 4(F^4)_{a_1 a_2 a_3 a_4}+4(F^4)_{a_1 a_2 a_4 a_3}+4(F^4)_{a_1 a_3 a_2 a_4}- (F^2)_{a_1 a_2}(F^2)_{a_3 a_4}\right. \right.\nonumber \\
&& \left.\left.- (F^2)_{a_1 a_3}(F^2)_{a_2 a_4}- (F^2)_{a_1 a_4}(F^2)_{a_2 a_3}\right] + {\rm~permutations~}a_2,a_3,a_4 \right\}
\end{eqnarray}
These terms give rise to contributions that are not gauge invariant since from
$p\rightarrow i\pa$ ordinary derivatives are produced. The correct 
covariantization of the result is obtained adding terms with one and two 
background connections $\mathbf{\Gamma}_{\a\ad}$ from the five- and six-point 
functions respectively. They are shown in Fig.2b, Fig.3a and Fig.3b.
By taking into account the contributions from these graphs we get the complete
covariantization \cite{noi} of the derivative expression in (\ref{eq:p^2}). By
futher taking into account the  properties of the trace operation and after some integrations by parts we obtain \\
\begin{minipage}[l]{1 \linewidth}
\begin{eqnarray} \label{eq:p^2finale}
&& \!\!\!\!\!\!\! \G_4^{(der)}= \nonumber \\
&& \frac{1}{20}\left(\frac{\pi^2}{12}\frac{1}{M^6}\right) \Tra \Big\{
(\DC F_{\m\n} \DC F_{\n\rho}F_{\rho\s}F_{\s\m}
+\DC F_{\m\n} \DC F_{\rho\s}F_{\n\rho}F_{\s\m}+ F_{\m\n}\DC F_{\rho\s} \DC F_{\n\rho}F_{\s\m})
 \nonumber \\
&&\left. -\frac{1}{4}(\DC F_{\m\n} \DC F_{\m\n}F_{\rho\s}F_{\rho\s}
+F_{\m\n}\DC F_{\m\n}\DC F_{\rho\s}F_{\rho\s}
+\DC F_{\m\n}\DC F_{\rho\s}F_{\m\n}F_{\rho\s})\right\} \nonumber \\
&& \nonumber \\
&& -\frac{1}{20}\left(\frac{\pi^2}{12}\frac{1}{M^6}\right) \Tra \left\{
-2\DC^2 F_{\m\n}F_{\n\rho}F_{\rho\s}F_{\s\m}
-4 \DC^2 F_{\m\n}F_{\rho\s}F_{\n\rho}F_{\s\m} \right. \nonumber \\
&& \left.~~~~~~~~~+ \DC^2 F_{\m\n}F_{\m\n}F_{\rho\s}F_{\rho\s} 
+\frac{1}{2}\DC^2 F_{\m\n}F_{\rho\s}F_{\m\n}F_{\rho\s} \right\}
\end{eqnarray}
\end{minipage}
\vskip 3mm
\noindent
where it is understood that the indices on the two $\nabla$ are to be 
contracted.

Let us observe that quite generally we can rewrite terms which contain  $\DC^2$ and four $F$'s  as $F^5$ contributions. Indeed, using the equations of motion $\nabla^\m F_{\m\n}=0$ and the Bianchi identities $ \nabla_{[\m} F_{\n\rho]}=0$,  we have the following identities (see \cite{noi})
\begin{eqnarray}\label{eq:D^2}
\DC^2 F_{\m\n}&=& 2i (F_{\m\s}F_{\s\n}-F_{\n\s}F_{\s\m})
\end{eqnarray}
Thus in order to study corrections to the symmetrized trace prescription one has to consistently take into account derivative contributions. In particular the above relation clearly shows that for the case of two covariant derivatives one has to consider not only the antisymmetrized products $\nabla_{[\m }\nabla_{\n]}$, as it was already noticed \cite{BRS}, but also the  symmetrized ones $\nabla_{(\m }\nabla_{\n)}$. 
Using (\ref{eq:D^2}), we can rewrite the last two lines of the 2-derivative result in (\ref{eq:p^2finale}) as $F^5$ and $F^2 F^3$ contributions.\\
Let us now consider the five-point function with two ${\bf W}$ and three 
${\bf {\Wb}}$ vertices as in Fig.2a. After completion of the $D$-algebra in 
the loop we obtain terms 
with five scalar propagators and a typical background dependence of the form 
\begin{equation}
\int d^4\theta~{\bf W}^\a(p_1,a_1){\bf W}_\a(p_2,a_2){\bf \Wb}^\ad(p_3,a_3)\Db_\ad{\bf \Wb}^\bd(p_4,a_4)
{\bf \Wb}_\bd(p_5,a_5)
\end{equation}
The final result at the leading order in the local expansion is:
\begin{eqnarray}
\label{eq:5puntifinale}
\G_{5}&=& -\frac{1}{32}\left(\frac{\pi^2}{12}\frac{1}{M^6}\right) ~ g(a_1,a_2,a_3,a_4,a_5)[-2 (F^5)_{a_1a_2a_3a_4a_5}-(F^5)_{a_1a_4a_2a_5a_3} 
\nonumber \\ \nonumber \\
&& +3(F^5)_{a_1a_4a_3a_2a_5}+1/2 (F^2)_{a_2a_4}(F^3)_{a_1a_3a_5}-1/2 (F^2)_{a_3a_4}(F^3)_{a_1a_2a_5}]
\end{eqnarray}
\begin{minipage}[r]{0.40 \linewidth}\begin{center}
\includegraphics[width=0.8 \linewidth]{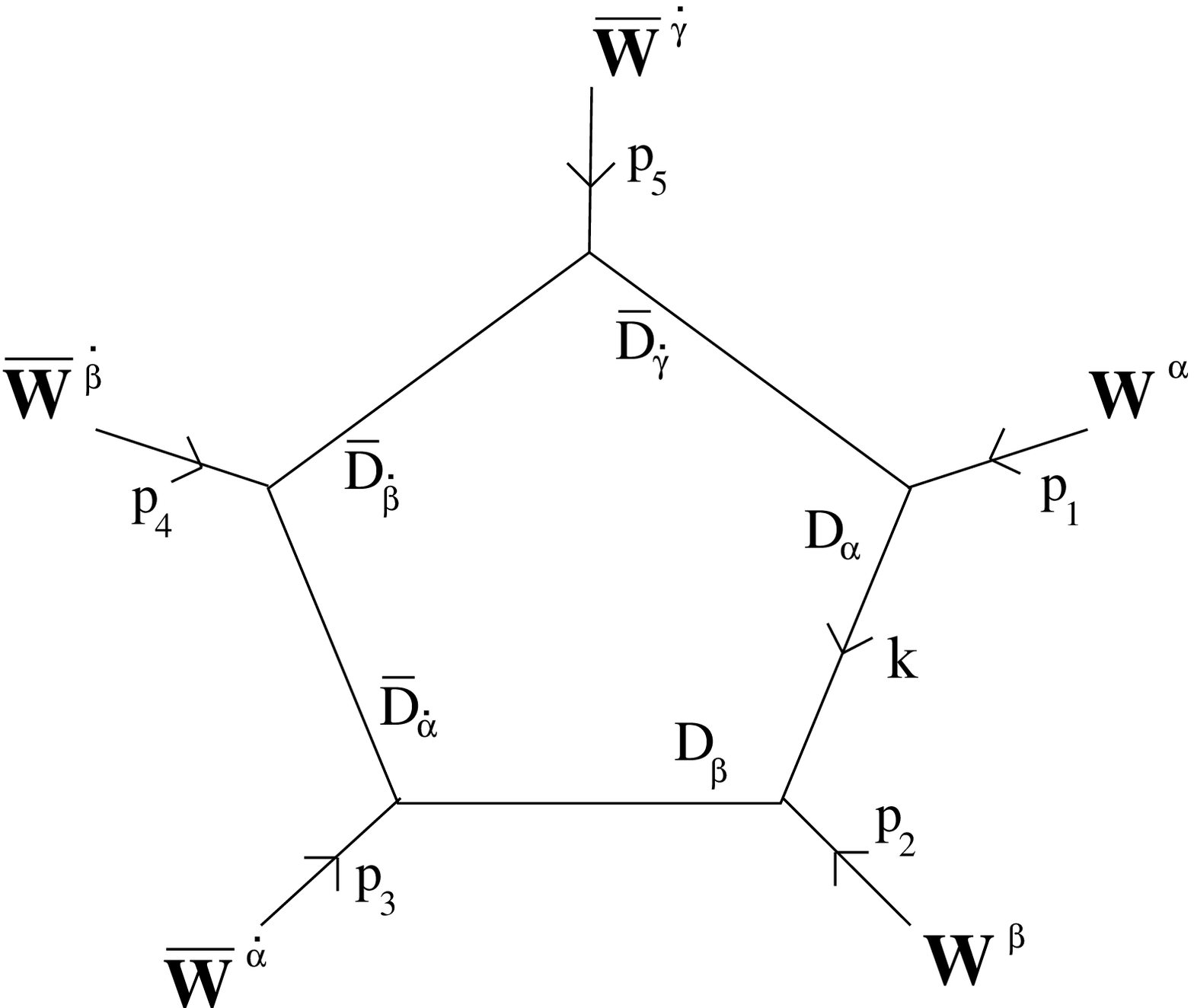} \\
a) \end{center}
\end{minipage}
\hskip 19.2mm
\begin{minipage}[l]{0.40 \linewidth}\begin{center}
\includegraphics[width=0.8 \linewidth]{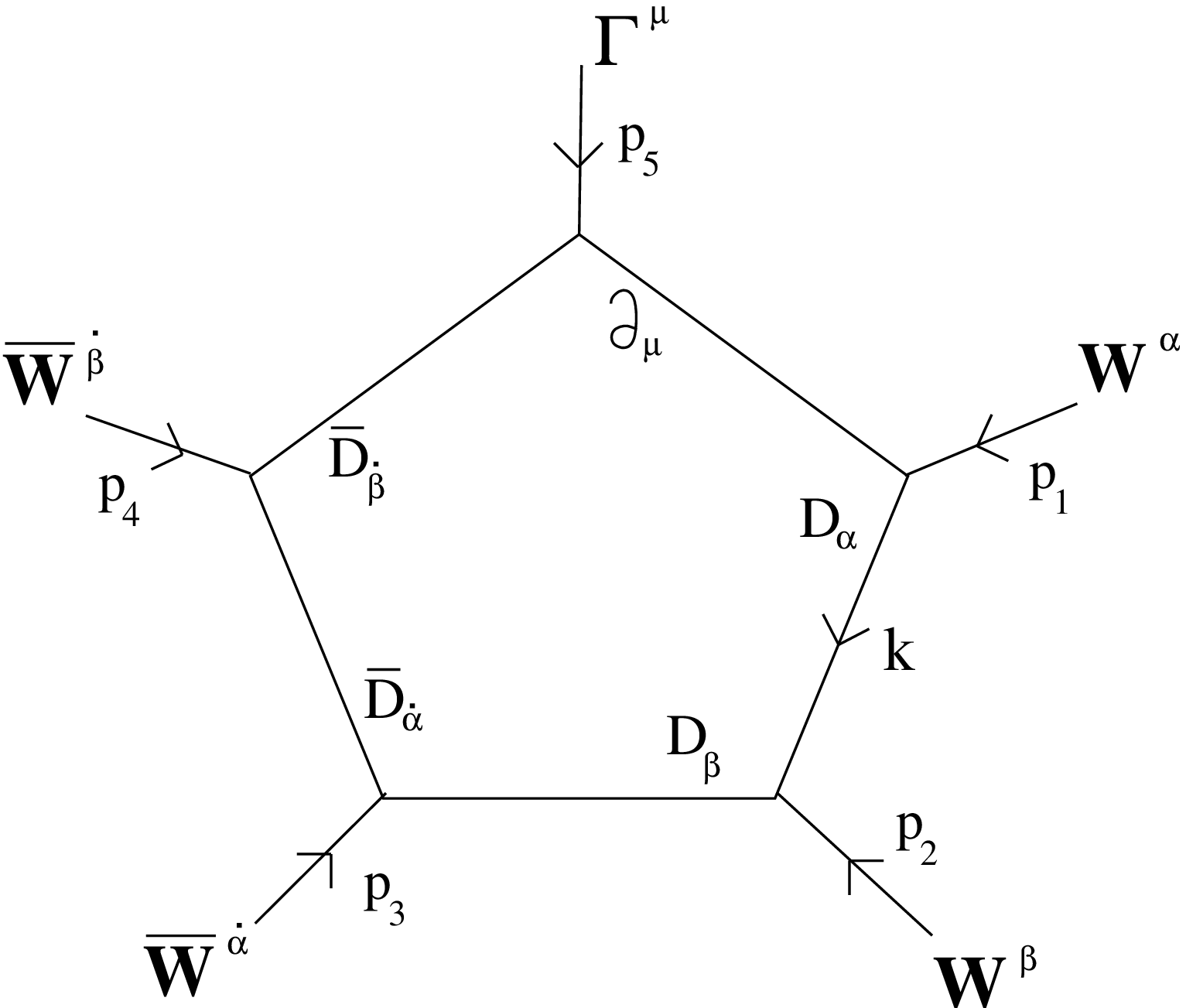} \\
b) \end{center}
\end{minipage}
\begin{center}
{\small {Figure 2: Five-point amplitudes}}
\end{center}
\hskip -4.0mm
\begin{minipage}[l]{0.45 \linewidth}\begin{center}
\includegraphics[width=0.7 \linewidth]{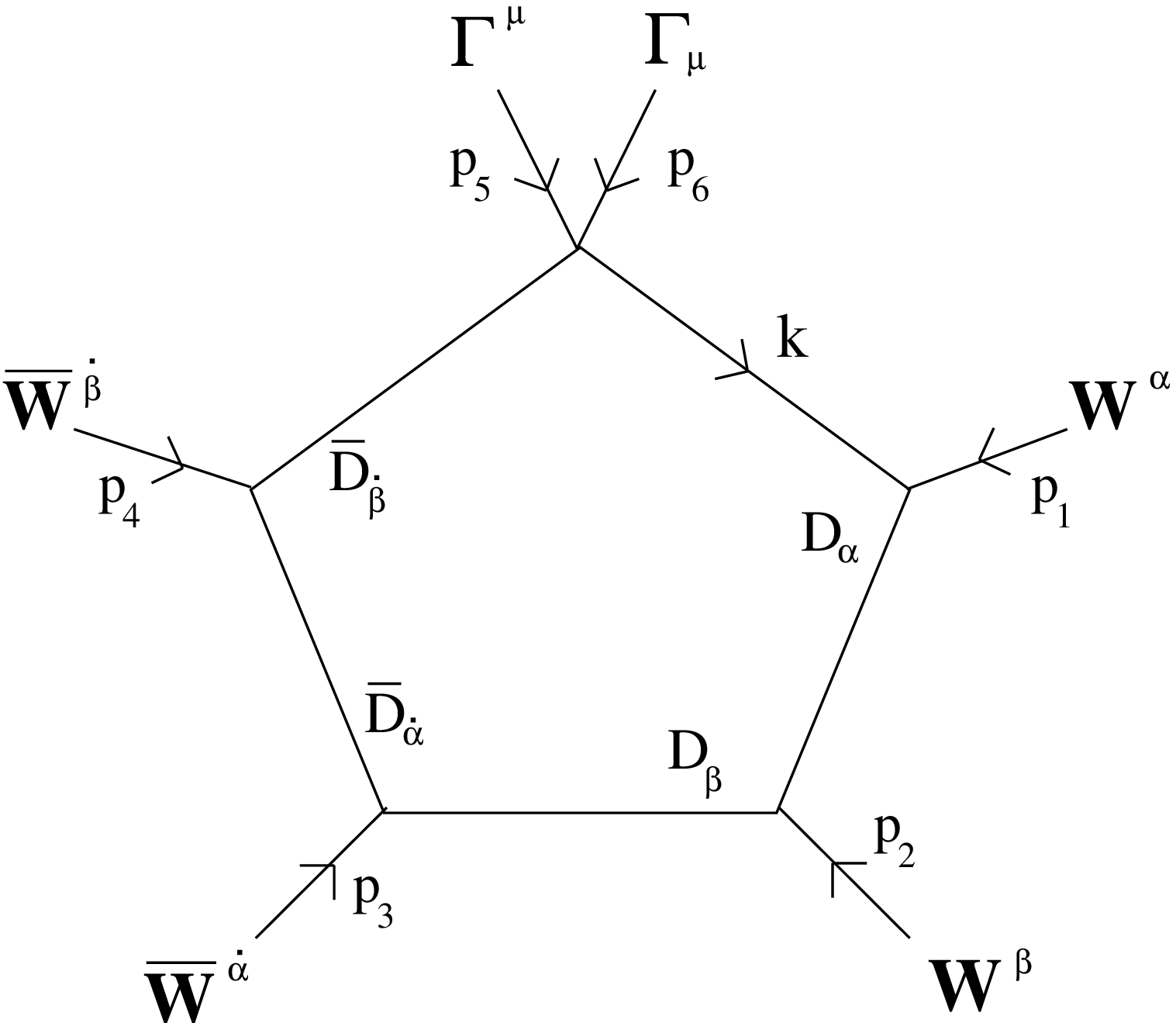}\\
a) \end{center}
\end{minipage}
\hskip 15.3mm
\begin{minipage}[r]{0.40 \linewidth}\begin{center}
\includegraphics[width=0.7 \linewidth]{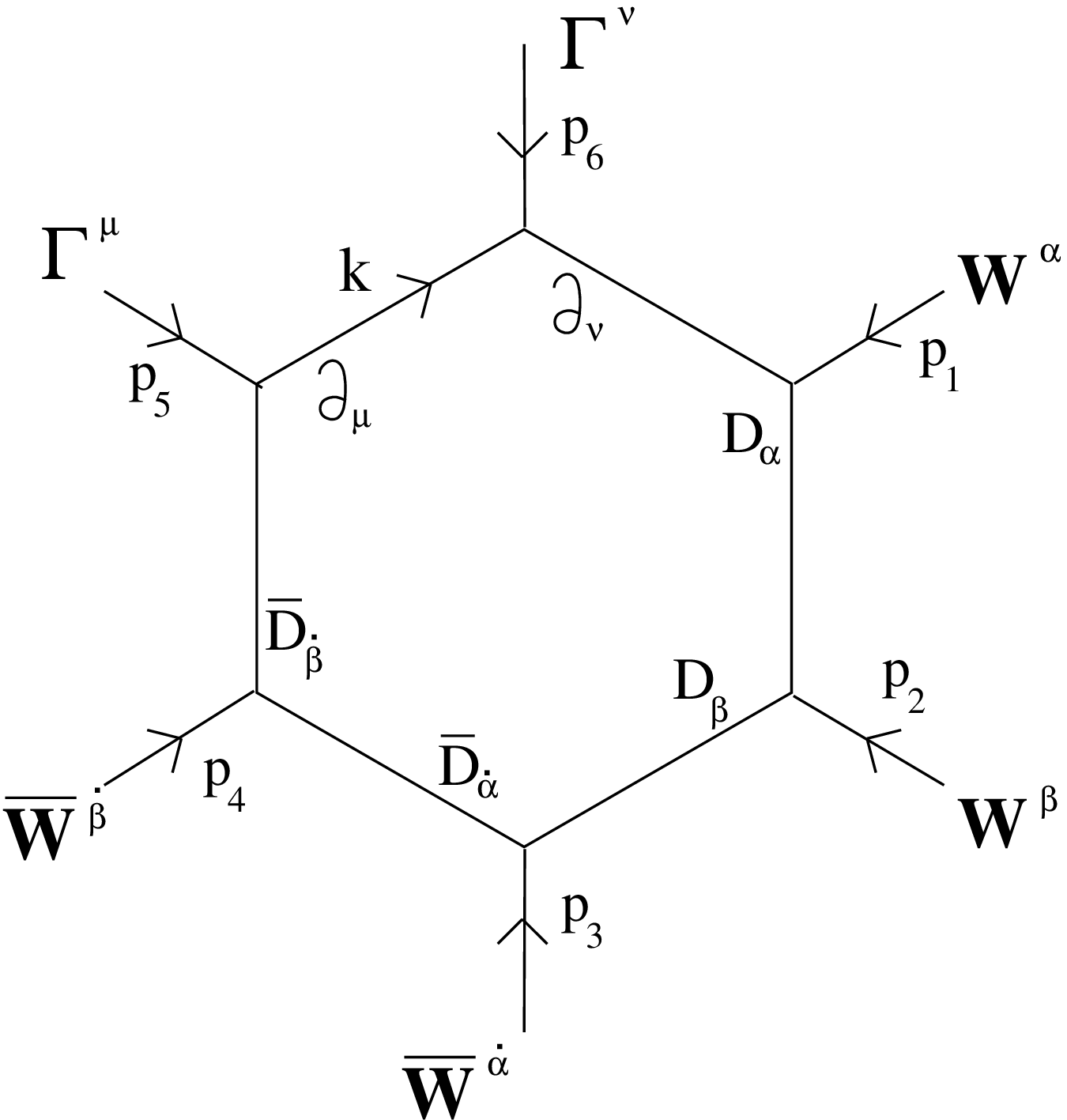} \\
b) \end{center}
\end{minipage}
\begin{center}
{\small {Figure 3: Six-point amplitudes}}
\end{center}
Let us now collect the results (\ref{eq:p^2finale}) and (\ref{eq:5puntifinale}); making use of the identities given in Appendix C of \cite{noi} we can eliminate $F^2F^3$ in favor of $F^5$ terms.
Our final result to the $M^{-6}$ order is:
\begin{eqnarray}
\label{eq:totale}
\G_{tot}&=& \frac{1}{20}\left(\frac{\pi^2}{12}\frac{1}{M^6}\right) \Tra \Big\{
(\DC F_{\m\n} \DC F_{\n\rho}F_{\rho\s}F_{\s\m}
+\DC F_{\m\n} \DC F_{\rho\s}F_{\n\rho}F_{\s\m}+ F_{\m\n}\DC F_{\rho\s} \DC F_{\n\rho}F_{\s\m})
 \nonumber \\
&&\left. -\frac{1}{4}(\DC F_{\m\n} \DC F_{\m\n}F_{\rho\s}F_{\rho\s}
+F_{\m\n}\DC F_{\m\n}\DC F_{\rho\s}F_{\rho\s}
+\DC F_{\m\n}\DC F_{\rho\s}F_{\m\n}F_{\rho\s})\right\} \nonumber \\
&& +\frac{1}{20}\left(\frac{\pi^2}{12}\frac{1}{M^6}\right)\Tra \left(
- F_{\m\n}F_{\n\rho}F_{\rho\s}F_{\s\t}F_{\t\m} 
-\frac{7}{2}F_{\m\n}F_{\n\rho}F_{\s\t}F_{\rho\s}F_{\t\m} 
 \right. \nonumber \\
&& \left.~~~~~~~~~~~~~~-\frac{3}{2}F_{\m\n}F_{\rho\s}F_{\t\m}F_{\n\rho}F_{\s\t} +2 F_{\m\n}F_{\rho\s}F_{\n\rho}F_{\t\m}F_{\s\t} \right)
\end{eqnarray}
Up to an overall numerical factor, the first two lines in (\ref{eq:totale}) 
reproduce the corresponding result in ref. \cite{kitazawa}, formula (3.3), 
obtained from an open superstring scattering amplitude.
Moreover they agree with the recent results recently obtained in \cite{Bil,KS}.
So, our first result is that, at the level of the four-point function, the 
supersymmetric Yang-Mills effective action exactly reproduces the structure 
of the non-abelian Born-Infeld theory, including the first derivative 
corrections. The terms $F^5$, which were also computed in \cite{kitazawa}, 
are more difficult to compare with ours: this is essentially due to the fact 
that the result quoted in \cite{kitazawa} is not written in a canonical form 
and moreover it requires additional symmetrizations which we are not clear 
how to interpret unambiguously. However, they do not agree with the result
found in \cite{KS}.\\
We would like to conclude summarizing our results.
We have considered the ${\cal N}=4$ supersymmetric Yang-Mills theory and 
computed at one-loop the four- and five-point functions with external vector 
fields. From the superfield result we have extracted the part of the bosonic 
components which contain the field strengths $F_{\m\n}$. These non-local 
one-loop contributions have been expanded in a low energy approximation and 
expressed as a sum of an infinite series of local terms. We have argued that 
these local expressions reproduce contributions to the non-abelian Born-Infeld 
action, if supersymmetry has to determine its structure. However, the discrepancy with the result found in \cite{KS} seems to indicate that supersymmetry is not enough at this order.


\vspace{0.2cm}

{\bf Acknowledgements}: the authors have
been partially supported by INFN, MURST, and the European 
Commission RTN program HPRN-CT-2000-00113 in which they are 
associated to the University of Torino. 

\end{document}